# Effects of annealing on the fluctuation conductivity and pseudogap in slightly doped HoBa$_2$Cu$_3$O$_{7-\delta}$ single crystals

A. L. Solovjov[1,2,3], L. V. Omelchenko[1], E. V. Petrenko[1], Yu. A. Kolesnichenko[1],
A. S. Kolesnik[1], S. Dzhumanov[4], and R. V. Vovk[3]

[1]*B. Verkin Institute for Low Temperatures Physics and Engineering of the National Academy of Sciences of Ukraine
Kharkiv 61103, Ukraine*

[2]*Institute for Low Temperatures and Structure Research, Polish Academy of Sciences, Wroclaw 50-422, Poland*

[3]*The Faculty of Physics, V. N. Karazin Kharkiv National University, Kharkiv 61022, Ukraine*

[4]*Institute of Nuclear Physics, Uzbek Academy of Sciences, Ulugbek, Tashkent 100214, Uzbekistan*
E-mail: solovjov@ilt.kharkov.ua

The effect of annealing at room temperature on the fluctuation conductivity (FLC) $\sigma'(T)$ and pseudogap (PG) $\Delta^*(T)$ in the basal *ab* plane of ReBa$_2$Cu$_3$O$_{7-\delta}$ (Re = Ho) single crystals with a lack of oxygen has been studied. It is shown that at all stages of annealing, the FLC near $T_c$ can be described by the Aslamazov–Larkin and Maki–Thompson fluctuation theories, demonstrating a 3D–2D crossover with increasing temperature. The crossover temperature $T_0$ was used to determine the coherence length along the *c* axis, $\xi_c(0) = (2.82 \pm 0.2)$ Å. At the intermediate stage of annealing, an anomalous increase in 2D FLC was revealed, which is associated with the influence of uncompensated magnetic moments in HoBa$_2$Cu$_3$O$_{7-\delta}$ (HoBCO): $\mu_{\text{eff, Ho}} = 9.7\mu_B$. For the quenched sample S1, the temperature dependence of the PG has a shape typical of single crystals with a large number of defects. However, $\Delta^*(T)$ has two small additional maxima at high temperature, which is a feature of HoBCO single crystals with pronounced twins and indicates the two-phase nature of the sample. Upon annealing, the shape of $\Delta^*(T)$ noticeably changes, very likely due to an increase in the magnetic interaction (sample S2). More important is the change in the slope of the data at high temperatures, which has become about 3.5 times steeper. The ordering of the oxygen distribution due to the diffusion process during annealing somewhat compensates for the influence of magnetic interaction. But the slope does not change (sample S3). Interestingly, the slope turns out to be the same as for FeAs-based superconductors, suggesting the possibility of the existence of spin density waves in HoBCO in the PG state. The comparison of the pseudogap parameter $\Delta^*(T)/\Delta^*_{\max}$ near $T_c$ with the Peters–Bauer theory revealed a slight increase in the density of local pairs $\langle n\uparrow n\downarrow \rangle$, which should explain the observed increase in $T_c$ by 9 K during annealing.

## 1. Introduction

In modern solid state physics, one of the urgent problems is the construction of a theory of high-temperature superconductors (HTSCs), which is necessary to elucidate the possibility of creating new superconductors with even higher, preferably room, critical temperatures $T_c$ of transition to the superconducting (SC) state. The solution of this problem is complicated by the lack of a clear understanding of the physics of internal interactions in such multicomponent compounds as HTSCs, in particular, the mechanism of SC pairing [1], which makes it possible to have a very high critical temperature of the SC transition [2]. It is believed that the answer to the question of SC pairing, as well as the possible role of the interplay between superconductivity and magnetism in the formation of paired fermions at temperatures above 100 K [2] in HTSCs, can be obtained by studying such an interesting phenomenon as a pseudogap (PG), which opens in cuprate HTSCs, type ReBa$_2$Cu$_3$O$_{7-\delta}$ (Re = Y, Ho, Gd, Pr) at temperature $T^* \gg T_c$. PG is a special state of matter, which is characterized by a reduced (but not to zero) density of electronic states at the Fermi level and a probable transformation of the Fermi surface below $T^*$ [1–5]. However, the physics of PG and its role in the formation of coupled electrons (fluctuating Cooper pairs) (FKPs) above $T_c$ are still unclear, despite the huge number of theoretical and experimental works devoted to this problem.

It has long been established that the behavior of HTSCs in the normal state goes far beyond the standard Fermi liquid approach [6–9]. As a result, a large number of non-Fermi liquid models [10–12] as well as marginal Fermi liquid models [13] have been proposed. All these models largely explain various specific aspects of the behavior of cuprates observed in the experiment. However, until now there is no unified theory that would be able to describe all the features of the behavior of HTSCs in the PG state. Recently, there has been a noticeable increase in interest in studying the problem of PG in HTSCs [3, 14–18]. Moreover, in addition to the already mentioned spin fluctuations [1, 7–9], spin density waves (SDW) [12, 17], charge ordering (CO) [1, 17], charge density waves (CDW) [15–18] as well as pair density waves (PDW) [19, 20] are proposed to explain the PG physics. We share a different approach to the problem of the PG appearance, which assumes the possibility of the formation of paired fermions, the so-called local pairs (LPs), in HTSCs below the PG opening temperature $T^* >> T_c$ [5, 21–23]. Moreover, it can be assumed that SDW, CDW, CO, and PDW can be considered as different possible mechanisms of LP pairing in cuprates in the PG state.

One of the most interesting materials for studying PG are $ReBa_2Cu_3O_{7-\delta}$ compounds, which is due to the possibility of wide variation of their composition by replacing yttrium with its isoelectronic analogues, or by changing the degree of oxygen nonstoichiometry. As is known, in the ground state all $ReBa_2Cu_3O_{7-\delta}$ are Mott dielectrics with a long-range antiferromagnetic (AF) order, in which the electron spins $S = 1/2$ are localized on copper ions $Cu^{2+}$ [6, 24]. The dielectric state is a consequence of strong electronic (Hubbard) correlations [3, 4] (and references therein). When charge carriers (holes) appear during doping, the long-range AF order is rapidly violated. However, as shown by neutron experiments with YBCO samples of different doping levels less than optimal, well-developed short-range order AF fluctuations [25] (and references therein) are retained in the normal metal phase of cuprates, which are observed up to very high doping levels ($T_c \sim 85$ K) [5].

Taking into account the presence of AF correlations in cuprates, of particular interest are compounds with partial replacement of yttrium (Y) by praseodymium (PrBCO with a magnetic moment $\mu_{eff} \approx 2\mu_B$) and especially with complete replacement by holmium, $HoBa_2Cu_3O_{7-\delta}$ (HoBCO), which has a magnetic moment $\mu_{eff} \approx 9.7\mu_B$, due to magnetic moment of pure Ho which is $\mu_{Ho} \approx 10.6\mu_B$ [5, 26]. However, PrBCO, being a magnetic dielectric, where all electrons are localized in the Fehrenbacher–Rice zone [27], quickly suppresses superconductivity in YBCO — the so-called "praseodymium anomaly". Whereas, HoBCO shows almost the same high $T_c$ values as YBCO.

Substitution of Y in single crystals of $ReBa_2Cu_3O_{7-\delta}$ (Re = Y, Ho, Dy, etc.) with Ho of a fairly large magnetic moment suggests a change in behavior of the system due to paramagnetic properties of $HoBa_2Cu_3O_{7-\delta}$ in its normal state [28]. Samples exhibiting oxygen nonstoichiometry are of special interest. In this state the redistribution of labile oxygen and structural relaxation take place simultaneously, thus significantly affecting electron transport parameters of the system [29, 30]. In this case rare alkaline earth metal ion can serve as a sensor, sensitive to local symmetry of its surroundings and charge density distribution, which can be affected by outside factors such as temperature [29], high pressure [30], or room-temperature annealing [31, 32].

To study the physical nature of the interaction between superconductivity and magnetism, single crystals were chosen, which have the advantage that their properties can noticeably change either during annealing of samples in an oxygen atmosphere due to an increase in the density of charge carriers $n_f$, or due to the formation of additional defects, as a result of rapid quenching from $\sim 600$ °C, followed by improving of their parameters by holding the samples in air at room temperature (the so-called room-temperature annealing) immediately after fabrication [31, 32]. The effect of annealing at room temperature (hereinafter simply annealing) on $T_c$, $n_f$ and the change in the lattice parameters of oxygen-deficient $ReBa_2Cu_3O_{7-\delta}$ (Re = Y, Ho) single crystals after their quenching from $T = 600$ °C is explained by the ordering of their structure due to the ordering of oxygen atoms in the $CuO_2$ planes without a noticeable change in the oxygen content in the sample [31, 32] (and references therein). Nevertheless, in spite of a number of studies on the relaxation processes in the 1–2–3 system during annealing [31–33] or, for example, under high pressure [34] (and references therein), many aspects, such as the charge transfer and the nature of the redistribution of the vacancy subsystem, still remain uncertain. The study of the influence of intrinsic magnetism Ho on the fluctuation conductivity (FLC) and PG in $HoBa_2Cu_3O_{7-\delta}$ single crystals upon annealing at room temperature is considered very promising for elucidating the mechanisms of the mutual influence of superconductivity and magnetism in HTSCs, which is important for the final elucidation of the nature of both PG and HTSC in general

This work is devoted to the study of the effect of annealing on the temperature dependences of the resistivity $\rho(T)$ of lightly doped $HoBa_2Cu_3O_{7-\delta}$ quenched single crystals, from which the temperature dependences of the excess conductivity $\sigma'(T)$ and pseudogap $\Delta^*(T)$ are calculated in the local pair model, as discussed in detail below. So far, no such studies have been carried out.

*Effects of annealing on the fluctuation conductivity and pseudogap in slightly doped* $HoBa_2Cu_3O_{7-\delta}$ *single crystals*

## 2. Experiment

$HoBa_2Cu_3O_{7-\delta}$ single crystals were grown with the solution-melt technique in a gold crucible as described elsewhere [33]. Rectangular crystals of about $1.9 \times 1.5 \times 0.2$ mm were selected to perform the resistivity measurements. The smallest parameter of the crystal corresponds to the *c* axis. A fully computerized setup utilizing the four-point probe technique with stabilized measuring current of up to 10 mA was used to measure the *ab* plane resistivity, $\rho_{ab}(T)$ [34]. Silver epoxy contacts were glued to the opposite sides of the crystal in order to produce a uniform current distribution in the central region where voltage probes in the form of parallel silver stripes were placed. Contact resistances below 1 Ω were obtained. Temperatures were measured with a Pt sensor having an accuracy of about 1 mK.

To reduce the oxygen content, the samples were annealed at 600 °C in ambient atmosphere for 24 h. After the annealing step, the crystals were quenched to room temperature within 2 min to 3 min, mounted on the holder and further cooled down to liquid-nitrogen temperature within 15 min. All measurements were conducted while warming up the samples. Experimental runs were carried at rates of about 0.1 K/min near $T_c$ and 0.5 K/min at $T >> T_c$.

In order to determine the room-temperature annealing effect on the electrical conductivity, after the first measurement of $\rho(T)$ (sample S1), the samples were kept at room temperature for 20 h, after which the measurements were repeated (sample S2). The last series of measurements was done after the room-temperature annealing of the samples for about 5 days (120 h) (sample S3), depending on completion of relaxation processes. Temperature dependencies of resistivity $\rho(T) = \rho_{ab}(T)$ for $HoBa_2Cu_3O_{6.65}$ single crystal with initial $T_c = 63.6$ K and $\delta \approx 0.35$ (sample S1) and measured after annealing during 20 h (sample S2) and 120 h (sample S3) are shown in Fig. 1. In the Insert to Fig. 1 shows a special approach to a more accurate determination of the PG opening temperature $T^*$ using the criterion $(\rho(T) - \rho_0)/aT = 1$, where $a = d\rho/dT$ [2].

Thus, we have three curves or actually three samples. The sample parameters are listed in the Tables. All curves have an expected S-shape with positive thermally activated bending, characteristic for slightly doped cuprates [23, 35]. However, above $T^* \approx 256.5$ K (S1), $\approx 238$ K (S2), and $\approx 235$ K (S3) $\rho(T)$ varies linearly with $T$ at rates $d\rho/dT =$ = 2.48, 1.91, and 1.57 μΩ·cm·K$^{-1}$ for S1, S2, and S3, respectively. The figure shows that after the first annealing, the resistivity rapidly decreases, but reaches saturation, when the annealing time increases up to 120 h [31].

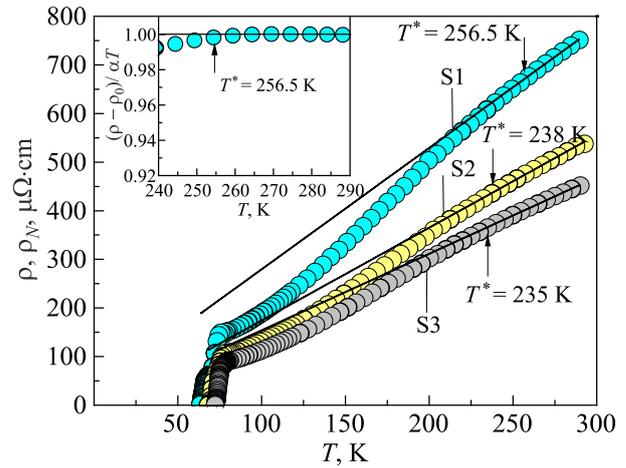

*Fig. 1.* (Color online) Temperature dependences of the resistivity $\rho_{ab}$ of single crystals of $HoBa_2Cu_3O_{6.65}$ for different stages of room temperature annealing: S1 — without annealing, S2 — annealing for 20 h, S3 — annealing for five days (120 h). The straight solid lines determine the normal-state resistivity $\rho_N$ (see the text). The Inset shows the method for determining $T^*$ using the criterion $(\rho(T) - \rho_0)/aT = 1$ [2].

At the same time, during annealing $T_c$ of the samples increases (Table 1), while $T^*$ decreases (Table 2) in good agreement with phase diagram of cuprates [1, 17, 35]. All dependences also show a clear trend towards saturation upon annealing (see the Tables). Interestingly, neither $\rho(T)$ nor $T_c$ practically increase with an increase in the annealing time above 120 h. Rapid cooling of crystals from about 600 °C (quenching) leads to the appearance of numerous additional disordered defects in the form of oxygen vacancies in $CuO_2$ planes [31] (and references there). It is these defects that are responsible for the increased resistivity of sample S1. It can be concluded that the decrease in $\rho(T)$ and the increase in $T_c$ observed during annealing are due to the ordering of labile oxygen in the $CuO_2$ planes due to diffusion but without an increase in the oxygen content.

A decrease in the slope $d\rho/dT$ indeed indicates the thermally activated character of the resistance relaxation process. The fact that the relaxation activation energy $\rho(T)$ coincides with the oxygen diffusion activation energy confirms the possibility of oxygen ordering in the planes. It has been shown that the oxygen diffusion, at least in YBCO single crystals, can proceed at an average rate of 150 Å per day [31].

Table 1. The resistivity and FLP parameters of the $HoBa_2Cu_3O_{6.65}$ single crystal

| Sample | $\rho(300 K)$, μΩ·cm | $\rho(100 K)$, μΩ·cm | $T_c$, K | $T_c^{mf}$, K | $T_G$, K | $T_0$, K | $T_{01}$, K | $\Delta \ln \sigma'$ | $d_{01}$, Å | $\xi_c(0)$, Å |
|---|---|---|---|---|---|---|---|---|---|---|
| S1 | 751.5 | 192.1 | 63.6 | 70.65 | 71.0 | 74.8 | 108.2 | 0.68 | 3.87 | 2.82 |
| S2 | 537.5 | 130.1 | 69.6 | 73.88 | 74.6 | 77.5 | 125.3 | 1.0 | 3.10 | 2.60 |
| S3 | 451.5 | 112.1 | 72.7 | 74.35 | 75.0 | 78.9 | 113.4 | 0.60 | 3.90 | 2.89 |

Table 2. The pseudogap parameters of the HoBa$_2$Cu$_3$O$_{6.65}$ single crystal

| Sample | $T^*$, K | $\alpha^*$ | $\varepsilon_{c0}^*$ | $C_{3D}$ | $A_4$ | $D^*$ | $\Delta^*(T_G)$, K | $\Delta^*(T_{max})$, K | $T_{max1}$, K | $T_{max2}$, K | $\Delta T_{max}$, K |
|---|---|---|---|---|---|---|---|---|---|---|---|
| S1 | 256.5 | 1.56 | 0.64 | 2.1 | 26 | 2.5 | 156.1 | 190.1 | 199.7 | 214.5 | 14.8 |
| S2 | 238 | 1.4 | 0.71 | 1.35 | 20 | 2.5 | 170.5 | 221.9 | 202.7 | 217.5 | 14.8 |
| S3 | 235 | 1.5 | 0.67 | 2.85 | 40 | 2.4 | 189.3 | 232.8 | 200.2 | 215.0 | 14.8 |

All these facts confirm the conclusion made. It appears that oxygen vacancies in slightly doped cuprates make the oxygen rearrangement easier to achieve.

Figure 2 shows the resistivity curves near $T_c$ for S1 (a) and S3 (b), respectively, in which all representative temperatures are designated. The observed stepwise resistive transition is characteristic of lightly doped HTSC single crystals, especially after quenching [31–34]. This is very likely due to the fact that rapid quenching from 600 °C also leads to a nonstoichiometric ratio of oxygen and vacancies, which, in turn, leads to the formation of separate phases in the sample. These phases are characterized by different oxygen content and its ordering and, accordingly, have different $T_c$ [31–33]. This is precisely what leads to the experimentally observed stepwise superconducting transitions (Fig. 2). The intersection of straight lines with the $x$ axis shows how $T_c$ was determined.

From Fig. 2 it follows that $T_c$ increases during annealing from 63.6 K (S1) to 72.7 K (S3) (Table 1), while the width of the resistive transition noticeably decreases by about 1.8 times. Accordingly, the two-step shape $\rho(T)$ practically disappears [Fig. 2(b)], which suggests the ordering of the crystal structure due to the redistribution of oxygen, as discussed above

The details of the superconducting transitions are best seen in the temperature dependences of the derivatives $d\rho(T)/dT$ obtained upon annealing at 20 °C, as was shown in Refs. 32, 36. In this case, the SC transitions are characterized by the low-temperature (LT) and high-temperature (HT) $d\rho(T)/dT$ maxima, indicating the presence of two superconducting phases with different $T_c$, as mentioned above. Upon annealing, the low-temperature peak shifts toward the high-temperature peak. After 120 h of annealing the LT peak becomes the highest and most uniform, and the HT peak is practically suppressed. Moreover, the distance between the peaks decreases as expected, which indicates a decrease in the width of the SC transition during annealing by about a factor of two (see Fig. 2) [32]. Interestingly, high hydrostatic pressure also reduces the resistance and increases $T_c$ of lightly doped HoBCO single crystals; however, in contrast to annealing, the width of the superconducting transition increases in this case [34]. This fact allows us to conclude that annealing, in contrast to pressure, creates a different mechanism of oxygen redistribution in a single crystal.

The effect of annealing on the parameters of lightly doped samples ReBCO (Re = Y, Ho) is usually interpreted by the ordering of oxygen atoms in the CuO$_2$ plane without changing the oxygen content in the sample (refer to [31, 32] and references therein), as mentioned above. However, an increase in $T_c$ during annealing can be associated either with the local ordering of oxygen [37] or with an increase in the density of charge carriers $n_f$ [38]. Another reason for the increase in $T_c$ can be a change in the parameters of the crystal cell, for example, change in Cu–O and Cu–Cu distances in the $ab$ plane [39]. In turn, the electrical resistivity decreases not only due to the ordering of oxygen atoms during the holding of the sample at room temperature after quenching from a high temperature [32]. But, it can also decrease due to an increase in $n_f$ or a decrease in defects as a result of the ordering of oxygen vacancies [31]. The possibility of increasing the charge carrier density $n_f$ upon annealing is discussed below.

Thus, we can conclude that, strictly speaking, many aspects of lightly doped quenched HoBCO single crystals, such as charge transfer and the nature of the redistribution of the vacancy subsystem, especially with allowance for the high Ho magnetism, still remain undetermined. We hoped to shed more light on this problem by studying the effect of annealing on FLC and PS in lightly doped quenched HoBCO single crystals.

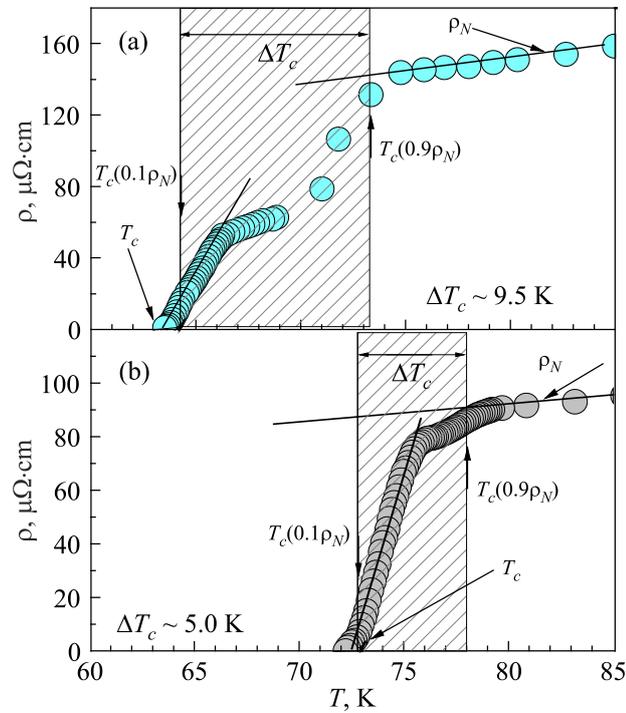

*Fig. 2.* (Color online) Resistive transitions of single crystals of HoBa$_2$Cu$_3$O$_{6.65}$: (a) without annealing, sample S1 (turquoise dots), (b) after annealing for five days (120 h), sample S3 (gray dots).

## 3. Results and discussion

### 3.1 Fluctuation conductivity

As clearly seen from Fig. 1, below the PG opening temperatures $T^* \gg T_c$ the resistivity curves of HoBa$_2$Cu$_3$O$_{6.65}$ single crystal deviate downward from linear dependencies at high temperatures. This resulting in appearance of the excess conductivity

$$\sigma'(T) = [\rho_N(T) - \rho(T)] / [\rho(T)\rho_N(T)], \quad (2)$$

as the difference between the experimentally measured resistivity $\rho(T)$ and the linear resistivity of the normal state $\rho_N(T) = \rho_0 + aT$, extrapolated towards low temperatures. Here, $\rho_0$ is the residual resistivity determined by extrapolating $\rho_N(T)$ towards 0 K, and $a = d\rho/dT$ is the slope of the linear straight line. This procedure of the normal state resistivity determination is widely used in literature (see [2, 5, 26, 40, 41] and references therein) and has been justified theoretically by the nearly antiferromagnetic Fermi liquid (NAFL) model [8]. The $T^*$ temperature is taken at the point where the experimental resistivity curve starts to downturn from the high-temperature linear behavior (Fig. 1). For a more accurate determination of $T^*$, the criterion $(\rho(T) - \rho_0)/aT = 1$ [2, 5, 40] was also used (see Insert in Fig. 1). Both approaches give practically the same $T^*$'s.

In the local pairs model, it is believed that PG and excess conductivity are due to the formation of LPs below $T^*$ [3, 5, 14, 22]. The properties of the LPs are determined by the coherence length along the $c$ axis $\xi_c(T) = \xi_c(0)[(T - T_c^{mf})/T_c^{mf}]^{-1/2} = \xi_c(0)\varepsilon^{-1/2}$ [42, 43], where $\varepsilon = (T - T_c^{mf})/T_c^{mf}$ is the reduced temperature and $T_c^{mf}$ is the mean-field critical temperature, which separates the range of SC fluctuations above $T_c$ from the region of critical fluctuations around $T_c$, where the SC order parameter $\Delta < kT$ [43, 44]. Hence, it is evident that the correct determination of $T_c^{mf}$ is decisive in FLC and PG analysis. It is well established [5, 23, 40, 45–47] that the small coherence length in combination with the quasi-layered structure of HTSCs leads to the formation of a noticeable, in comparison with conventional superconductors, range of SC fluctuations, $\Delta T_{fl}$, above $T_c$. Near $T_c$, where $\xi_c(T) > d = 11.68$ Å ($d$ is the unit cell size along the $c$ axis [48]), the LPs are very large and interact throughout the entire volume of the sample, forming a 3D state. In this case the FLC is always described by the Aslamazov–Larkin equation for any 3D system [49]:

$$\sigma'_{3DAL} = C_{3D} \frac{e^2}{32h\xi_c(0)} \varepsilon^{-1/2}, \quad (3)$$

where $C_{3D}$ is a numerical factor used to fit the data by the theory [23, 44]. This means that the conventional 3D FLC is realized in HTSCs as $T \to T_c$ [43, 50]. From Eq. (3), one can easy obtain $\sigma'^{-2} \sim \varepsilon \sim (T - T_c^{mf})/T_c^{mf}$. Obviously, $\sigma'^{-2} = 0$, when $T = T_c^{mf}$. This way of $T_c^{mf}$ determination was proposed by Beasley [44] and substantiated in various FLC experiments [5, 26, 50]. Moreover, with the correct choice of $T_c^{mf}$, the data in the three-dimensional fluctuation region near $T_c$ are always approximated by Eq. (3).

Figure 3 shows the $\sigma'^{-2}$ vs $T$ plot for samples S1 [(a) turquoise dots] and S3 [(b) gray dots]. The interception of the extrapolated linear $\sigma'^{-2}$ with $T$ axis determines $T_c^{mf} = 70.65$ K (S1) and $= 74.35$ K (S3). S2 (not shown) demonstrates some intermediate $\sigma'^{-2}$ on $T$ dependence with $T_c^{mf} = 73.88$ K given in Table 1. It should be noted, that the revealed dependences of $\sigma'^{-2}$ on $T$ differ markedly. When the measurements were carried out immediately after quenching (S1), the separation between the LT and HT phases is pronounced (see Fig. 2), and S1 exhibit a dependence $\sigma'^{-2}$ on $T$ characteristic of most HTSCs. But in this case, the LT phase, which is clearly seen in Fig. 2(a) has virtually no effect on the definition of $T_c^{mf}$. However, unlike Fig. 2(b), after annealing, both the LT and HT phases plotted in these coordinates become quite pronounced. But, fortunately and somewhat surprisingly, the approximation of both phases by straight lines gives the same $T_c^{mf}$.

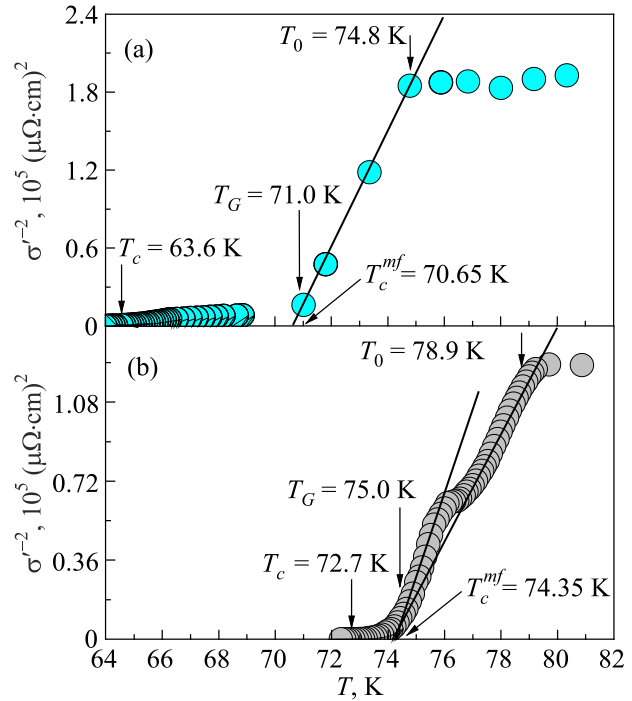

*Fig. 3.* (Color online) Temperature dependences of the inverse square of the excess conductivity $\sigma'^{-2}(T)$ for sample S1 [(a), turquoise dots] and S3 [(b), gray dots)]. The interception of the extrapolated linear $\sigma'^{-2}$ with $T$-axis determines $T_c^{mf}$. Also shown are $T_c$, the Ginsburg temperature $T_G$ and 3D–2D crossover temperature $T_0$.

Above the crossover temperature the data deviates right from the line suggesting the 2D Maki–Thompson (MT) [42, 51, 52] fluctuation contribution into FLC [23]. Obviously, at the crossover temperature $T_0 \sim \varepsilon_0$ the coherence length $\xi_c(T_0) = \xi_c(0)\varepsilon_0^{-1/2}$ is expected to amount to $d$ [9, 37] which yields

$$\xi_c(0) = d\sqrt{\varepsilon_0} \qquad (4)$$

and allows the possibility of $\xi_c(0)$ determination. $\xi_c(0)$ is one of the important parameters of the PG analysis.

Figure 4 shows the $\ln\sigma'$ vs $\ln\varepsilon$: (a) S1 (turquoise dots), (b) S2 (yellow dots), and (c) S3 (gray dots) in comparison with the fluctuation theories. As expected, all samples demonstrate fairly good agreement with the AL theory near $T_c$. For example, above the Ginzburg temperature $T_G > T_c^{mf}$ ($\ln\varepsilon_G = -5.3$) (refer to Fig. 3), down to which the mean-field theory works [50], and up to $T_0 = 74.8$ K ($\ln\varepsilon_0 = -2.84$) the data for sample S1 are well extrapolated by the 3D fluctuation term (3) of the AL theory, Fig. 4(a), solid red line with a slope –1/2) with $\xi_c(0) = (2.82 \pm 0.02)$ Å determined by Eq. (4) and $C_{3D} = 2.1$ (see Table 2). It should be noted that the same value of $\xi_c(0)$ was determined for a FeAs-based superconductor ErFeAsO$_{0.85}$F$_{0.15}$ [53]. Samples S2 and S3 behave similarly near $T_c$ (Fig. 4).

Above $T_0$, the data deviate sharply upwards from the AL theory. This is due to the fact that at $T \geq T_0$, where $\xi_c(T) < d$, the three-dimensional regime ends. However, it is still $\xi_c(T) > d_{01}$, which is the distance between conducting planes CuO$_2$ [48], and $\xi_c(T)$ connects the planes with the Josephson interaction. This is a 2D fluctuation regime, which is described by the MT term of the Hikani–Larkin (HL) theory [42]:

$$\sigma'_{MT} = C_{2D}\frac{e^2}{8d\hbar}\cdot\frac{1}{1-\alpha/\delta}\cdot\ln\left((\delta/\alpha)\cdot\frac{1+\alpha+\sqrt{1+2\alpha}}{1+\delta+\sqrt{1+2\delta}}\right)\varepsilon^{-1}. \qquad (5)$$

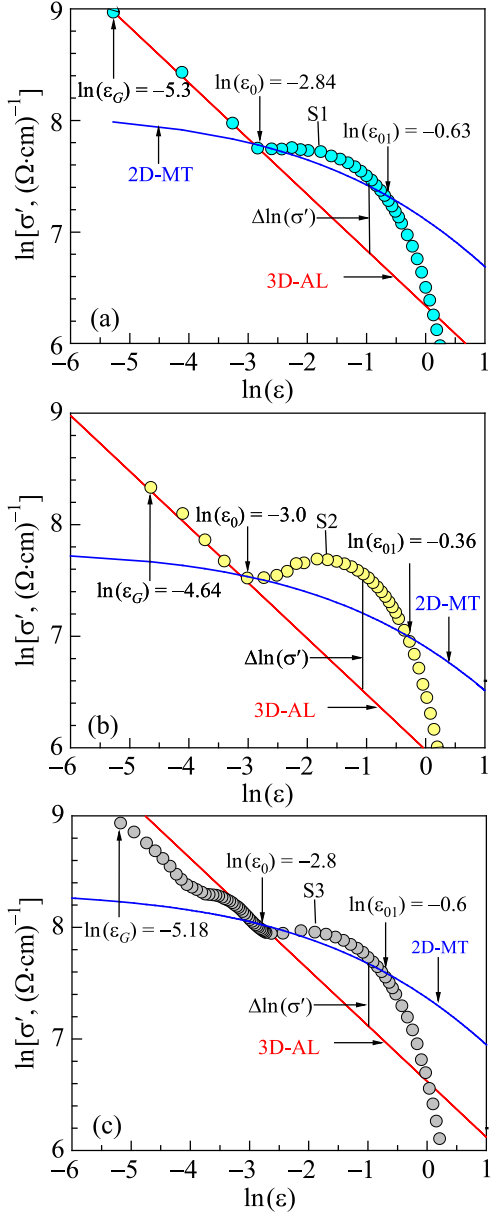

*Fig. 4.* (Color online) $\ln\sigma'$ vs $\ln\varepsilon$: (a) S1 (turquoise dots), (b) S2 (yellow dots), and (c) S3 (gray dots) compared with fluctuation theories: 3D-AL — red lines; 2D-MT — blue curves. $\Delta\ln(\sigma')$ designates the maximal deviation of the data from the extrapolated 3D-AL lines.

These are the blue solid curves in the figure. In Eq. (5) $\alpha = 2[\xi_c(0)/d]^2\varepsilon^{-1}$ is a coupling parameter,

$$\delta = \beta\frac{16}{\pi h}\left[\frac{\xi_c(0)}{d}\right]^2 k_B T\tau_\varphi \qquad (6)$$

is the pair-breaking parameter, and $\tau_\varphi$ that is defined by equation

$$\tau_\varphi\beta T = \pi h/8k_B\varepsilon_0 = A/\varepsilon_0 \qquad (7)$$

is the phase relaxation time, where $A = 2.998\cdot10^{12}$ s·K. The factor $\beta = 1.203$ ($l/\xi_{ab}$), where $l$ is the mean-free path and $\xi_{ab}(0)$ is the coherence length in the $ab$ plane considering the clean limit approach $l > \xi$, which is always takes place in HTSCs [5, 6, 26, 40, 43, 50].

Above $T_{01}$, indicated on all graphs as $\ln\varepsilon_{01}$, the data of all samples deviate definitively downward from the theory (Fig. 4). Thus, $T_{01}$ limits the range of SC fluctuations. In this range, fluctuating pairs behave much like ordinary Cooper pairs, but without long-range ordering (the so-called short-range phase correlations [5, 22, 23, 44]). Above $T_{01}$, $\xi_c(T) < d_{01}$ and LPs are confined within the CuO$_2$ planes, which are no longer connected by any correlation interaction. Thus, it is clear that $\xi_c(T_{01}) = d_{01}$. To estimate $d_{01}$, we use the condition $\xi_c(0) = d\sqrt{\varepsilon_0} = d_{01}\sqrt{\varepsilon_{01}} = (2.82 \pm 0.02)$ Å (S1). Since $d = c = 11.68$ Å and $\ln\varepsilon_{01} \approx -0.63$ ($\varepsilon_{01} = 0.532$ and $T_{01} \approx 108.2$ K), we obtain:

$d_{01} = d\sqrt{\varepsilon_0}\sqrt{\varepsilon_{01}} = (3.87 \pm 0.05)$ Å for S1 in good agreement with results of the structural studies [48]. Having carried out a similar analysis for other samples, we obtain the values of $\xi_c(0)$ and $d_{01}$ for S2 and S3 (Table 1). Interestingly, in contrast to S1 and S2, in the case of S3, both the LT and HT phases are clearly visible on the plot of $\ln \sigma'$ vs $\ln \varepsilon$ below $T_0$, but both fully follow the AL theory. However, to keep the logic with S1, we determined $T_0$ and other parameters from the high temperature phase.

Strictly speaking, extrapolation of 2D-MT data above $T_0$ is not entirely successful. This is due to the fact that above $T_0$ there is a sharp increase in data, leading to the appearance of enhanced 2D fluctuations. As a result, the maximal deviation of the data above the extrapolated 3D-AL line $\Delta \ln \sigma' = 0.68$ obtained for S1 [Fig. 4(a)] is approximately 3.4 times greater than that observed for YBCO, where magnetism is not expected [5, 23, 47]. This enhanced behavior of 2D fluctuation is typical of FeSe-based superconductors such as ErFeAsO$_{0.85}$F$_{0.15}$ [53] and SmFeAsO$_{0.85}$ [54], as well as superconductors with magnetic impurities such as YBCO–PrBCO superlattices [50], suggesting a noticeable influence of own HoBCO magnetism in our case.

The highest value $\Delta \ln \sigma' = 1.0$, which is approximately 5 times greater than that observed for YBCO, was obtained for S2 [Fig. 4(b)]. In this case, fitting the data by the MT theory is completely impossible. However, to provide a more informative analysis, we used the found fitting parameters ($\xi_c(0)$, $\varepsilon_0$, $\varepsilon_{01}$) to derive theoretical 2D-MT curve that would intersect the red 3D-AL line at $\ln \varepsilon_0$ (corresponding temperature $T_0$) and the 2D-data at $\ln \varepsilon_{01}$ (corresponding temperature $T_{01}$) [Fig. 4(b)]. We emphasize that, despite the unsatisfactory description of the 2D data, all temperatures $T_{01}$ found in this way (indicated in the figure as $\ln \varepsilon_{01}$) clearly correspond to the minima on the temperature dependences of the PG parameter $\Delta^*(T)$, which follows from the theory (see Fig. 7), thereby confirming the correctness of our analysis. The enhanced 2D fluctuations found for S2 are reminiscent of these observed for magnetic superconductors such as Dy$_{0.6}$Y$_{0.4}$Rh$_{3.85}$Ru$_{0.15}$B$_4$, which have an intrinsic magnetic moment $\mu \approx 6.2\mu_B$ per Dy$^{3+}$ ion [55]. Taking all above facts into account, we can conclude that the observed anomalous 2D fluctuations in sample S2 is most likely caused by the noncompensated magnetic moments of Ho, which is thought to be responsible for interplay between magnetic interaction and superconductivity [5, 26]. However, since the resistivity noticeably decreases upon annealing, it can be concluded that magnetic interaction does not strongly affect the rate of charge carrier scattering in HoBCO single crystals. At the same time, S2 has the lowest values of $\xi_c(0)$, $d_{01}$ (Table 1) and unexpectedly $C_{3D} = 1.35$, confirming a strong influence of oxygen diffusion on the sample structure [31, 32]. Recall that the smaller the $C_{3D}$, the smaller the effect of defects in the sample, which is directly related to the decrease in resistivity.

In turn, after five days of annealing, S3 demonstrates the lowest value $\Delta \ln \sigma' = 0.6$ and the shape of $\ln \sigma'$ vs $\ln \varepsilon$ resembling S1, except for the low temperature 3D-AL region [Fig. 4(c)]. This indicates the final ordering of defects and the crystal structure as a whole, which leads to the lowest resistivity (Fig. 1 and Table 1). It can also be assumed that ordered oxygen somehow shielded the influence of Ho magnetic moments. However, despite the supposed ordering of defects and oxygen, the values of $\xi_c(0)$, $d_{01}$ (Table 1) and $C_{3D} = 2.85$ are practically the same as in S1, which indicates a nonmonotonic, in contrast to the resistivity, effect of defects and magnetism on the FLC of the sample upon annealing. We expected to obtain confirmation or refutation of this conclusion by analyzing the change in the temperature dependence of the pseudogap during annealing.

### 3.2. Pseudogap analysis

As mentioned above, the number of the different non-Fermi-liquid models proposed to explain the physics of PG is quite large. However, a very large number of models raises doubts about their correctness. In addition, none of the mentioned models gives explicitly the temperature dependence of PG, which could be verified experimentally. Clearly, to attain information about the pseudogap we need an equation which specifies a whole experimental curve, from $T_G$ up to $T^*$, and contains the PG parameter $\Delta^*(T)$ in an explicit form. The issue was resolved within the framework of our LP model [23, 56], in which such an equation was proposed for $\sigma'(\varepsilon)$:

$$\sigma'(T) = A_4 \frac{e^2 \left(1 - \frac{T}{T^*}\right)\exp\left(-\frac{\Delta^*}{T}\right)}{16\hbar\xi_c(0)\sqrt{2\varepsilon_{c0}}\sinh\left(2\frac{\varepsilon}{\varepsilon^*_{c0}}\right)}, \quad (8)$$

where, for a correct description of the experiment, the dynamics of pair formation $(1 - T/T^*)$ and pair breaking $(\exp(-\Delta^*/T))$ above $T_c$ are taken into account. Solving Eq. (8) for the pseudogap $\Delta^*(T)$ one can readily obtain:

$$\Delta^*(T) = T \ln\left[A_4\left(1-\frac{T}{T^*}\right)\frac{1}{\sigma'(\varepsilon)}\frac{e^2}{16\hbar\xi_c(0)}\frac{1}{\sqrt{2\varepsilon_{c0}}\sinh\left(2\frac{\varepsilon}{\varepsilon^*_{c0}}\right)}\right]. \quad (9)$$

Here $\sigma'(\varepsilon)$ is the experimentally measured excess conductivity over the whole temperature interval from $T^*$ down to $T_G$, and $A_4$ is a numerical factor that has the meaning of the $C$ factor in the fluctuation conductivity theory [42–44, 51, 56]. All other parameters, including the coherence length along the $c$ axis, $\xi_c(0)$, [Eq. (4)] and the theoretical parameter $\varepsilon^*_{c0}$ [57], directly come from the experiment [5, 23, 50, 54–56]. To find $\varepsilon^*_{c0}$ we use the experimental fact that in some temperature range above $T_{01}$, namely $\ln \varepsilon_{c01} < \ln \varepsilon < \ln \varepsilon_{c02}$ (Fig. 5) or accordingly $\varepsilon_{c01} < \varepsilon < \varepsilon_{c02}$ (Insert in Fig. 5), $\sigma'^{-1} \sim \exp(\varepsilon)$ [23, 56, 57]. As a result $\ln(\sigma'^{-1})$ is a linear function of $\varepsilon$ with a slope $\alpha^* = 1.56$ which determines parameter $\varepsilon^*_{c0} = 1/\alpha^* = 0.64$ for S1 (Insert in Fig. 5). To find $A_4$, we calculate $\sigma'(\varepsilon)$ from $T^*$ and down to $T_G$ using Eq. (8) and fit experiment in the range of 3D-AL fluctuations near $T_c$ (Fig. 5, red curve) where $\ln \sigma'$ on $\ln \varepsilon$ is a linear function of the reduced temperature $\varepsilon$ with a slope $\lambda = -1/2$. Besides, $\Delta^*(T_G) = \Delta(0)$ is assumed [58, 59]. To estimate $\Delta^*(T_G)$, which we use in Eq. (8), we plot $\ln \sigma'$ as a function of $1/T$ for S1 (Fig. 6) and S3 (Insert to Fig. 6). After annealing (Insert to Fig. 6) the approximation, as expected, looks better, due to the ordering of defects. In this case the slope of the theoretical curve [Eq. (8)] turns out to be very sensitive to the value of $\Delta^*(T_G)$ [23, 50, 56]. For sample S1 the best fit is obtained when $D^* = 2\Delta^*(T_G)/k_B T_c = 5$ and $D^* = 4.8$ for S3 (Table 2). Note, that $D^* = 5$ is the typical value for cuprates [5, 60].

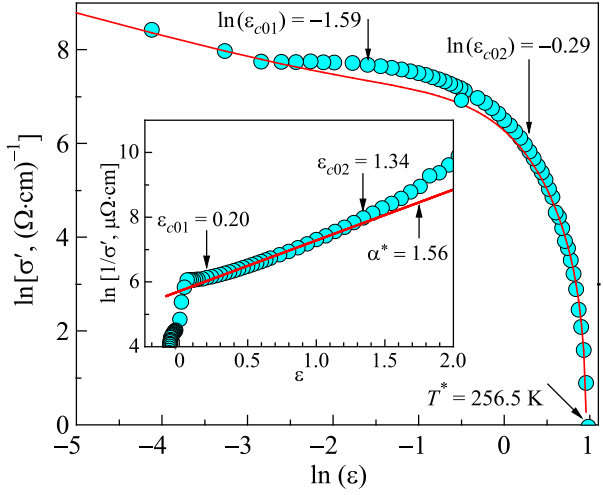

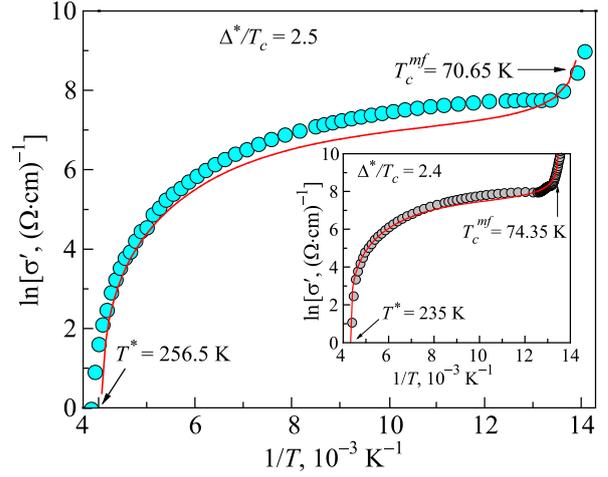

*Fig. 5.* (Color online) $\ln \sigma'$ vs $\ln \varepsilon$ for S1 (turquoise dots) plotted in the whole temperature range from $T^*$ down to $T_G$. The solid red curve is a fit to the data with Eq. (8). Insert: $\ln \sigma^{-1}$ as a function of $\varepsilon$. Solid line indicates the linear part of the curve between $\varepsilon_{c01} = 0.20$ and $\varepsilon_{c02} = 1.34$. Corresponding $\ln \varepsilon_{c01} = -1.59$ and $\ln \varepsilon_{c02} = -0.29$ are marked by arrows in the main panel. The slope $\alpha^* = 1.6$ determines the parameter $\varepsilon^*_{c0} = 1/\alpha^* = 0.64$ (Table 2).

*Fig. 6.* (Color online) $\ln \sigma$ vs $1/T$ for S1 (turquoise dots) and S3 (Insert, gray dots) plotted in the whole temperature range from $T^*$ down to $T_G$. The red solid curves are fits to the data with Eq. (8). The best fit is obtained when Eq. (8) is calculated with $D^* = 2\Delta^*(T_G)/k_B T_c = 5$ for S1 and $D^* = 4.8$ for S3 (Table 2).

Having determined all the necessary parameters (refer to Tables 1, 2) we succeeded to plot the temperature dependences PG, $\Delta^*(T)$ for all stages of annealing. Fig. 7(a) (turquoise dots) displays $\Delta^*(T)$ for S1 calculated using Eq. (9) with the following set of parameters derived from the experiment within the LP model: $T^* = 256.5$ K, $T_c^{mf} = 70.65$ K, $\xi_c(0) = 2.82$ Å, $\varepsilon^*_{c0} = 0.64$, $A_4 = 26$. The resulting form of $\Delta^*(T)$ with a high-temperature maximum at $T_{max} = 239.5$ K, followed by a section of the linear dependence $\Delta^*(T)$ with a moderate positive slope $\alpha_{max} = 1.16 \pm 0.01$, is typical of a lightly doped HTSC single crystals, containing various defects, including tweens [61] (and references therein).

This confirms our assumption made above about numerous defects in the quenched crystal. The low-temperature behavior of $\Delta^*(T)$ in S1 with minimum at $T_{01}$, maximum at about $T_0$ and final minimum at $T_G$ is also characteristic of all HTSCs (see Fig. 12 in [50]). The range of SC fluctuations $\Delta T_{fl} = T_{01} - T_G = 37.2$ K is large but comparable with $\Delta T_{fl}$ obtained for slightly doped YBCO single crystals with TBs [61]. The only peculiarities are two small maxima at $T_{max1} = 199.7$ K and $T_{max2} = 214.5$ K which is a feature of the PG behavior found only on HoBCO single crystals [26, 34].

Dependences $\Delta^*(T)$ constructed for the samples S2 and S3 with the corresponding sets of parameters given in the Tables, are displayed in Figs. 7(b) and 7(c), respectively. It can be seen that the shape of $\Delta^*(T)$ noticeably changed upon annealing. The linear section with a moderate positive slope disappeared, but maxima at $T_{max1}$ and $T_{max2}$ became much more pronounced. It is very tempting to attribute these maxima to two phases with different $T_c$ observed at resistive transitions (Figs. 2 and 3). But upon annealing, the separation into two phases gradually disappears, whereas, despite a significant change in all parameters of the sample during annealing, the distance between these maxima remains constant at $\Delta T_{max} = 14.8$ K (Table 2). There have also been attempts to relate these maxima to a process of the so-called ascending diffusion, which is believed to increase the separation of charge carriers between tweens and TBS [34] (and references therein). But the ascending diffusion also changes during the annealing, whereas the distance between these maxima remains constant, as discussed above. We believe that these maxima are somehow connected with enhanced magnetism of HoBCO, but, strictly speaking, the appearance of these unusual maxima is still in question.

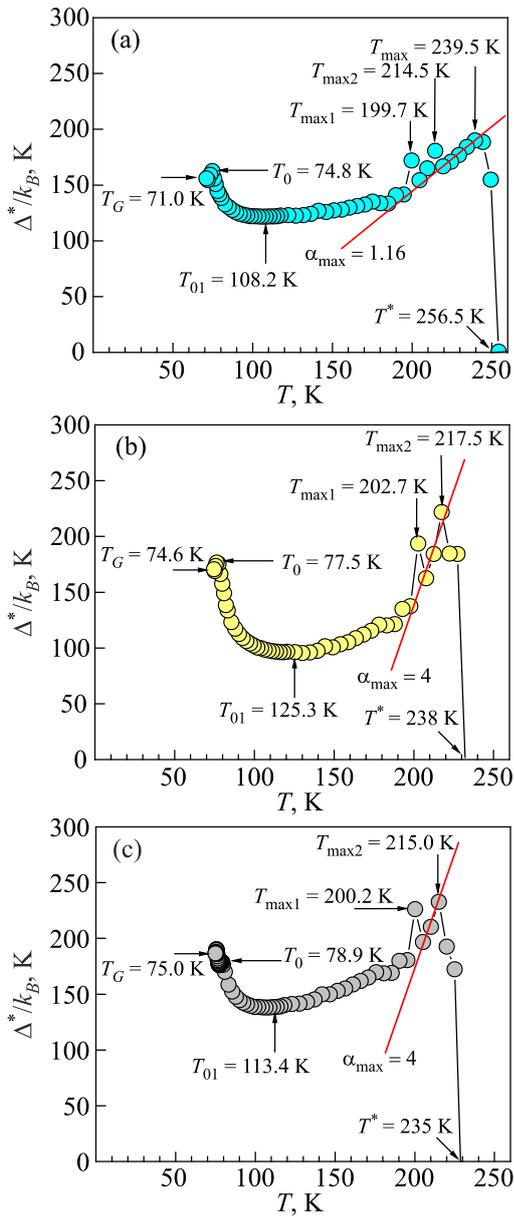

*Fig. 7.* (Color online) Temperature dependences of pseudogap $\Delta^*(T)$ (a) S1 (turquoise dots), (b) S2 (yellow dots), and (c) S3 (gray dots), analyzed with Eq. (9). All characteristic temperatures are marked with arrows. The red lines designate the data slope at high temperatures, which is equal for S2 and S3. Solid black lines are a guide for the eyes.

As seen in Fig. 7(b), S2 demonstrates a rather specific $\Delta^*(T)$ with pronounced wide minimum, as expected, at $T_{01} \approx 125.3$ K. This leads to an anomalously large range of SC fluctuations, $\Delta T_{fl} = T_{01} - T_G$, about 50 K above $T_c$. In addition, the two maxima look more pronounced and shifted slightly towards higher temperatures. And, more importantly, the data slope at high temperature is about 3.5 times steeper than that of the S1. Taking into account the results of the study of the FLC [Fig. 4(b)], we associate this form of $\Delta^*(T)$ with an increased magnetism of the uncompensated magnetic moments Ho. In turn, S3 [Fig. 7(c)] demonstrates $\Delta^*(T)$ characteristic of HoBCO single crystals with ordered defects [26, 34], which allows us to conclude that after five days of annealing, oxygen diffusion almost ceased. Indeed, the minimum at $T_{01} = 113.4$ K is noticeably smaller, and the high-temperature maxima are not as pronounced as for S2. However, the range of SC fluctuation $\Delta T_{fl} = T_{01} - T_G \approx 38$ K and the temperatures of these maxima in this case are almost the same as for the quenched sample S1 [Fig. 7(a)]. This allows us to draw the following conclusion that ordered oxygen somewhat shields the magnetic interaction in the crystal. On the other hand, in general, the shape of the dependence $\Delta^*(T)$ differs markedly from the shape of S1. The value of $\Delta^*(T_G)$, which gradually increases during annealing, reaches the maximum value of 189.3 K for S3 (Table 2). In addition, the minimum at $T_{01}$ is still quite pronounced and, more importantly, the slope of the data at high temperature $\alpha_{max} = 4.0$, marked in the figure by the red line, is the same as for S2.

Here we would like to emphasize that the same data slope at high temperatures is observed for all magnetic superconductors, including FeAs-based compounds (see [50], Fig. 11). This is confirmed by Fig. 8, where we compare our data for sample S3 with results obtained for the highly magnetic superlattice 7YBCO×14PrBCO (sample SL3) from Ref. 50. It can be seen that the slope at high $T$ is exactly the same. Moreover, the shape of both curves below $T_{01}$ is also almost the same. Interestingly, in 1111 FeAs-based superconductors the maximum corresponds to the structural transition from a tetragonal to an orthorhombic phase. Accordingly, the temperature of data deviation from the linear dependence corresponds to transition to the AF state of spin–density–waves (SDW) [62–65]. Thus, an important conclusion can be drawn that the AF interaction of the SDW type should take place in lightly doped HoBCO single crystals below the $T_{max1}$ (Fig. 7) due to the large intrinsic magnetism of Ho. This interaction is believed to be responsible for the formation of both PG and SDW state in such compounds. The possibility of SDW state in lightly doped YBCO compounds is discussed in Ref. 17. Thus, returning to the question of possible models of SC pairing in HTSCs, we can assume that the SDW model is the most probable one, at least for HTSCs with a strong magnetic interaction.

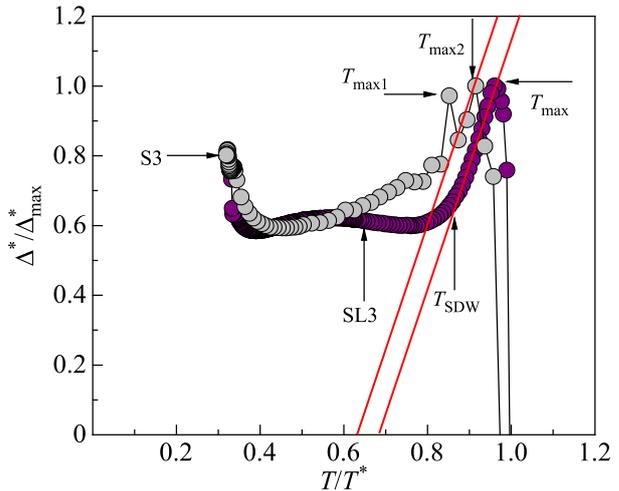

*Fig. 8.* (Color online) $\Delta^*(T)/\Delta^*_{max}$ as a function of $T/T^*$ for studied single crystals of HoBa$_2$Cu$_3$O$_{6.65}$ after annealing for five days (120 h), sample S3, and superlattice 7YBCO×14PrBCO, sample SL3. All characteristic temperatures are marked with arrows. The red lines designate the slope of the data at high temperatures, which is the same for both samples. Solid black lines are a guide for the eyes.

To clarify the question of a possible increase in the density of charge carriers in a crystal during annealing, we compare the pseudogap parameter $\Delta^*(T)/\Delta^*_{max}$ of samples S1, S2, and S3 near $T_c$ with the Peters–Bauer (PB) theory [3] (Fig. 9). In [3], the temperature dependences of the local pairs density $\langle n_\uparrow n_\downarrow \rangle$ in HTSCs were theoretically calculated within the framework of the three-dimensional attractive Hubbard model for different temperatures $T/W$, interactions $U/W$, and filling factor, where $U$ is the activation energy and $W$ is the band width. The shape of $\Delta^*(T)$ for all cuprates, with a maximum near $T_0$ followed by a minimum at $T_G$ [50] (see Fig. 12 in [50]) and Fig. 7 in [66]), resembles the shape of theoretical $\langle n_\uparrow n_\downarrow \rangle$ curves at low $T/W$ and $U/W$ [3]. This fact should justify such an approach.

To carry out the analysis, we combine the measured values of $\Delta^*(T)/\Delta^*_{max}$ for S1 at $T_G$ with the minimum, and at $T_0$ with the maximum of each theoretical curve calculated at different values of $U/W$, thus achieving the best agreement between the experiment and theory in the widest possible temperature range. It is important that the fitting factors found for S1 are also used for the other two samples [67, 68]. The fitting results for all three samples are shown in Fig. 9. The best fit for S1 near $T_c$ is obtained with $U/W = 0.2$ curve indicating that in this case $\langle n_\uparrow n_\downarrow \rangle \approx 0.3$, which is a typical value for various HTSCs [67, 68]. Further, it was taken into account that $\Delta^*(T_G)/\Delta^*_{max} = 0.82$ for S1, where $\Delta^*_{max}$ is taken at $T_{max} = 239.5$ K, which is clearly seen in Fig. 7(a). Unfortunately, due to the specific form of $\Delta^*(T)$ found for S2 and S3, leading to ambiguity in the definition of $T_{max}$, it was not possible to obtain reasonable values of $\Delta^*(T_G)/\Delta^*_{max}$ in this case. Therefore, we could not compare them with that found for S1 in order to obtain the corresponding fitting coefficients, as we did in our previous works [67, 68].

However, both $T_G$ and $\Delta^*(T_G)$ are clearly defined for all the studied samples (Fig. 7). Moreover, $\Delta^*(T_G)$ noticeably increases upon annealing (Table 2), most likely due to an increase in the charge carrier density $n_f$. This fact suggests that $n_f$ must be proportional to the value of $\Delta^*(T_G)$ [5, 23]. Taking into account that found for S1 $\Delta^*(T_G)$ corresponds to $\langle n_\uparrow n_\downarrow \rangle = 0.30195$, the simple algebra yields:

$$\langle n_\uparrow n_\downarrow \rangle (S2) = [\Delta^*(T_G, S2)/\Delta^*(T_G, S1)]$$

$$\times \langle n_\uparrow n_\downarrow \rangle (S1) = (170.5 \times 0.30195)/156.1 \approx 0.33, \text{ and}$$

$$\langle n_\uparrow n_\downarrow \rangle (S3) = = [\Delta^*(T_G, S3)/\Delta^*(T_G, S1)]$$

$$\times \langle n_\uparrow n_\downarrow \rangle (S1) = (189.3 \times 0.30195)/156.1 \approx 0.366,$$

which gives the corresponding curves at the figure. This means that the density of charge carriers in HoBCO single crystals somewhat increases due to oxygen diffusion during annealing, as it was assumed in Refs. 69, 70. It is very likely that the observed slight increase in $n_f$ is quite sufficient to explain the observed increase in $T_c$ by ~ 9 K. This may also be responsible to some extent for the observed decrease in $\rho(T)$ (Table 1).

It is also worth noting that the best agreement with the PB theory among HTSCs in a wide temperature range was obtained for non-twinned optimally doped YBCO single crystals, naturally, without any magnetism [67]. In the present case the data noticeably deviate downward from the theoretical curves with increasing temperature, which is most likely due to the enhancement of the magnetic interaction in HoBCO. This conclusion is confirmed by the following results. The quenched sample S1 shows the smallest deviation, since the magnetic moments Ho are considered to be randomly distributed due to multiple defects. The sample S2 shows the largest deviation, as a result of influence of the uncompensated magnetic moments, as discussed above. In the case of sample S3, the magnetic interaction is somehow compensated by the ordering of the distribution of oxygen and crystal structure defects [31, 32]. As a result, despite the rather complicated shape of the $\Delta^*(T_G)/\Delta^*_{max}$ curve at low $T$, the deviation from the theory is expectedly moderate (Fig. 9). Moreover, above $(T/W, T/T^*) \approx 0.25$, the experimental data deviate upward from the theory, which confirms a fundamentally different mechanism of magnetic interaction in the HoBCO single crystal after five days of annealing at room temperature.

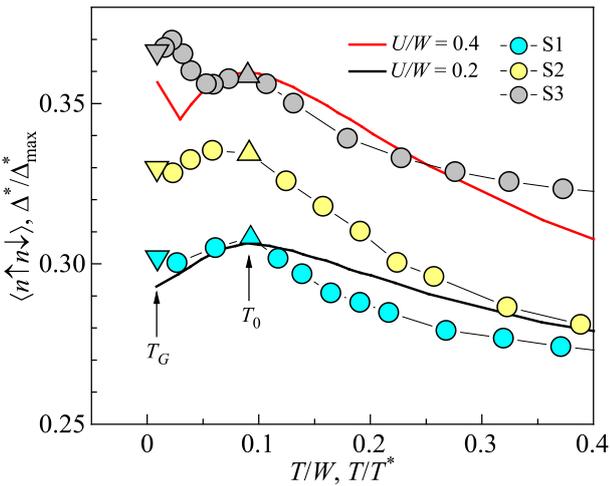

*Fig. 9.* (Color online) Curves of $\Delta^*/\Delta^*_{max}$ as functions $T/T^*$ for samples S1 (turquoise dots), S2 (yellow dots), and S3 (gray dots) in comparison with the theoretical curves of as functions of $T/W$, at the corresponding interaction values $U/W$: 0.2 (black curve), 0.4 (red curve). The arrows indicate the temperatures $T_0$ (▲) and $T_G$ (▼).

**Conclusion**

The magnitude and temperature dependence of fluctuation conductivity and pseudogap $\Delta^*(T)$ in lightly doped $HoBa_2Cu_3O_{7-\delta}$ single crystals rapidly quenched from 600 °C were studied for the first time at different stages of annealing at room temperature. During annealing, a significant decrease in the resistance of the samples, the width of resistive transitions, and an increase in $T_c$ were observed. These observations are consistent with the processes of the oxygen diffusion and structural relaxation in the volume of experimental samples, leading to the appearance of phase separation. However, in addition to the expected change in oxygen distribution, several new interesting results were revealed.

At all stages of annealing, the FLC near $T_c$ is well described by the 3D Aslamazov–Larkin fluctuation theory. However, at the intermediate stage of annealing (sample S2), an anomalous increase in 2D FLC was revealed, which is associated with the influence of uncompensated magnetic moments in $HoBa_2Cu_3O_{7-\delta}$, since $\mu_{eff, Ho} = 9.7 \mu_B$. As a result, in this case, the 2D Maki–Thompson fluctuation theory failed to describe the data. However, after five days of annealing, the 2D-MT fit improved, since the magnetic interaction is somehow compensated by the ordering of the distribution of oxygen and crystal structure defects.

For the quenched sample S1, the temperature dependence of the PG has a shape typical of single crystals with a large number of defects. However, $\Delta^*(T)$ has two small additional maxima at high temperature, which is a feature of $HoBa_2Cu_3O_{7-\delta}$ single crystals with pronounced twins and indicates the two-phase nature of the sample. Upon annealing, the shape of $\Delta^*(T)$ noticeably changes, very likely due to an increase in the magnetic interaction (sample S2). The two additional peaks became more pronounced. But, more important is the change in the slope $\alpha_{max}$ of the data at high temperatures, which has become about 3.5 times steeper. The ordering of the oxygen distribution due to the diffusion process during annealing somewhat compensates for the influence of magnetic interaction. But the slope does not change (sample S3). Moreover, the slope turns out to be the same as for FeAs-based superconductors, suggesting the possibility of the existence of spin density waves in $HoBa_2Cu_3O_{7-\delta}$ in the PG state. The comparison of the pseudogap parameter $\Delta^*(T_G)/\Delta^*_{max}$ near $T_c$ with the Peters–Bauer theory revealed a slight increase in the density of local pairs $<n\uparrow n\downarrow>$, which should explain the observed increase in $T_c$ by 9 K during annealing. Interestingly, despite the rather complicated shape of the $\Delta^*(T_G)/\Delta^*_{max}$ curve at low $T$ (sample S3), the deviation from the PB theory is expectedly moderate (Fig. 9). Moreover, above $(T/W, T/T^*) \approx 0.25$, the experimental data deviate upward from the theory, which confirms a fundamentally different, compare with S1 and S2, mechanism of magnetic interaction in the HoBCO single crystal after five days of annealing. Thus, the studies of FLC and PG turned out to be very informative and made it possible to obtain new results, which, in turn, are definitely a consequence of the expected rearrangement of the oxygen distribution and the defect structure during annealing at room temperature.

**Acknowledgments**

We thank support from the National Academy of Sciences of Ukraine through Young Scientists Grant No. 1/N-2021 (L. V. O. and E. V. P.). We acknowledge support from the Ministry of Innovative Development of the Republic of Uzbekistan through Grant No. F-FA-2021-433 (A. L. S., S. D., and R. V. V.). A. L. S. also thanks the Division of Low Temperatures and Superconductivity, INTiBS Wroclaw, Poland, for their hospitality.

______

___

Вплив відпалу на флуктуаційну провідність та псевдощілину у слаболегованих монокристалах $HoBa_2Cu_3O_{7-\delta}$

A. L. Solovjov, L. V. Omelchenko, E. V. Petrenko,
Yu. A. Kolesnichenko, A. S. Kolesnik, S. Dzhumanov,
R. V. Vovk

Вивчено вплив відпалу при кімнатній температурі на флуктуаційну провідність (ФЛП) σ′(*T*) і псевдощілину (ПЩ) Δ*(*T*) у базисній площині *ab* монокристалів $ReBa_2Cu_3O_{7-\delta}$ (Re = Ho) з нестачею кисню. Показано, що на всіх етапах відпалу ФЛП поблизу $T_c$ можна описати флуктуаційними теоріями Асламазова–Ларкіна та Макі–Томпсона, де спостерігається 3D–2D кросовер із підвищенням температури. За температурою кросовера $T_0$ визначено довжину когерентності вздовж осі *c* — $\xi_c(0) = (2{,}82 \pm 0{,}2)$ Å. На проміжному етапі відпалу виявлено аномальне зростання 2D ФЛП, що пов'язано з впливом некомпенсованих магнітних моментів у $HoBa_2Cu_3O_{7-\delta}$ (HoBCO): $\mu_{eff, Ho} = 9{,}7\mu_B$. Для загартованого зразка S1 температурна залежність ПЩ має форму, типову для монокристалів з великою кількістю дефектів. Проте Δ*(*T*) має два невеликі додаткові максимуми при високих температурах, що є особливістю монокристалів HoBCO з вираженими двійниками та вказує на двофазність зразка. Під час відпалу форма Δ*(*T*) помітно змінюється, ймовірно, за рахунок збільшення магнітної взаємодії (зразок S2). Більш важливою є зміна нахилу даних при високих температурах, який став приблизно в 3,5 рази крутішим. Упорядкування розподілу кисню за рахунок процесу дифузії під час відпалу дещо компенсує вплив магнітної взаємодії, проте нахил не змінюється (зразок S3). Цікаво, що нахил виявляється таким же, як і для надпровідників на основі FeAs, що свідчить про можливість існування хвиль спінової щільності в HoBCO в ПЩ стані. Порівняння псевдощілинного параметра Δ*(*T*)/Δ*$_{max}$ поблизу $T_c$ з теорією Пітерса–Бауера виявило незначне збільшення щільності локальних пар $<n_\uparrow n_\downarrow>$, що має пояснювати спостережене підвищення $T_c$ на 9 K під час відпалу.

Ключові слова: флуктуаційна провідність, псевдощілина, надлишкова провідність, відпал, монокристали HoBCO.